\documentclass[conference]{IEEEtran}

\usepackage{cite}
\usepackage{amsmath,amssymb,amsfonts}
\usepackage{algorithmic}
\usepackage{graphicx}
\usepackage{textcomp}
\usepackage{xcolor}
\usepackage{multirow}
\usepackage{booktabs}
\def\BibTeX{{\rm B\kern-.05em{\sc i\kern-.025em b}\kern-.08em
    T\kern-.1667em\lower.7ex\hbox{E}\kern-.125emX}}
\begin{document}

\title{\textbf{A New Transformer-Based Approach for Audio-Based Kinship Verification and a New Uncontrolled Mandarin Kinship Speech Dataset}}

\makeatletter
\renewcommand\footnoterule{%
  \kern-3\p@
  \hrule\@width.4\columnwidth
  \kern2.6\p@}
\makeatother
\IEEEoverridecommandlockouts 

\author{
    \IEEEauthorblockN{
        Qiyang Sun$^{1,*}$,
        Langqing Zhang$^{1,*,\dagger}$,
        Yupei Li$^{1}$,
        Björn W. Schuller$^{1,2}$
    }
    \IEEEauthorblockA{
        $^{1}$Department of Computing, Imperial College London, London, United Kingdom \\
        $^{2}$Chair of Health Informatics, TUM University Hospital, Munich, Germany \\
        \small{\textsf{\{q.sun23, langqing.zhang24, yupei.li22, bjoern.schuller\}@imperial.ac.uk, }}
    }
    \thanks{$^{*}$Equal contribution.}
    \thanks{$^{\dagger}$Corresponding author.}
}

\maketitle

\begin{abstract}
Kinship verification is a task involving determining whether two individuals share a first-order kin relation. To tackle this task, we propose CONVTRAP-TN, a new architecture for audio-based kinship verification, and conduct an ablation study. We additionally analyze the proposed model from the perspective of explainable AI (XAI), which is rarely, if at all, done in existing literature on audio-based kinship verification. Our architecture targets the issue of signal sparsity in kinship verification through attention mechanisms. To the best of our knowledge, we are the first to apply a transformer-based architecture to the task of audio-based kinship verification. Furthermore, we also collect a custom speech dataset, ARKIN, which accurately reflects everyday recording conditions. We do this because only a few speech datasets with kinship labels currently exist, all of which either source extremely noisy in-the-wild data from the internet, or instruct speakers to record in specific environments. These settings fail to reflect real-world scenarios where users record on personal devices under unrestrained conditions. Additionally, we perform a series of preliminary baseline experiments on the collected dataset, including speaker verification and recognition, speech recognition, age estimation, and kinship verification, as well as cross-dataset kinship verification experiments to show that existing methods are not robust across datasets.

\end{abstract}

\begin{IEEEkeywords}
Audio Kinship Verification, Transformer, Deep Metric Learning, Kinship Dataset, Cross-Corpus Evaluation
\end{IEEEkeywords}

\section{Introduction}
	\label{sec:introduction}
	Kinship verification involves deciding whether a kinship relation (first-order unless specified) exists between two individuals by analyzing signals collected from them. It is an important field of research with a broad range of applications, including searching for lost family members, forensics, home security, and authentication \cite{nih_kinship_review, talkin_dataset}. The kinship verification task is considerably more challenging than speaker-level tasks such as speaker verification \cite{kanav_dataset}. This can be remedied by incorporating data of multiple modalities, as seen in the large performance improvement of models trained on multiple modalities compared to models trained on a single modality \cite{kanav_dataset, talkin_dataset, fiw-mm_dataset}. Nevertheless, this also means that if the system does not have access to data of multiple modalities, performance severely degrades, which can easily be the case in real-world applications. Therefore, it is important to develop robust methods for various modalities individually \cite{sun2025explainable}. While visual kinship datasets are abundant \cite{cornell_kinface_dataset, kinfacew_dataset, fiw_dataset, uva-nemo_smile_dataset, kfvw_dataset}, audio-based counterparts remain scarce \cite{kanav_dataset, talkin_family_dataset, fiw-mm_dataset}.
	
	Furthermore, any finite dataset inevitably contains biases \cite{biases}, which negatively impact the ability of a model to stay invariant with respect to demographic and environmental features. This is a problem even in visual datasets \cite{bias_in_face_analysis,bias_in_visual_data}, which are much more abundant than audio. Researchers have attempted to mitigate this problem by applying explicit techniques to address dataset bias. For example, on a visual face dataset, we can use age progression to perform data augmentation \cite{mitigating_face_bias}, or on a speech dataset, convert audio samples to an age-standardized domain by performing age domain conversion \cite{age_domain_conversion}. Such efforts to mitigate algorithmic bias align with the core principles of Friendly AI (FAI) that remain beneficial to humanity in the future \cite{sun2026towards}. However, many features, such as the environment in which data is collected, cannot be explicitly modeled. For the task of audio-based kinship verification specifically, considering the limited number of publicly available audio kinship datasets, it is evident that robust audio-based kinship verification systems that can maintain performance across datasets are much harder to develop compared to systems with access to visual or multimodal data. To our best knowledge, almost all existing literature on audio-based kinship verification do not perform cross-dataset evaluation, which may be simply due to the limited number of datasets available. It is also acknowledged in literature that more robust voice models are needed \cite{robust_spk_rec_w_limited_data, robust_speaker_verification, robust_spk_ver_using_tts}, as they can in turn help improve fusion models that make use of audio data \cite{talkin_family_dataset}.
	
	On top of that, all publicly available audio kinship datasets either consist of purely in-the-wild data from public figures appearing in sources such as movies and YouTube videos \cite{kanav_dataset, talkin_dataset, fiw-mm_dataset}, or of data collected under specific environments with limited noise, such as a studio or a quiet room \cite{talkin_family_dataset}. Neither of these settings reflects average end-users interacting with audio recording systems under realistic conditions. Moreover, the existing body of audio data available for kinship verification consists primarily of English speech, which hinders progress in developing language-invariant models. Therefore, we construct a new, everyday-condition-oriented, purely audio dataset containing speech in Mandarin, with complete kinship, age, and gender labels, called Audio under Realistic conditions for KINship verification (ARKIN). We try to preserve natural recording conditions of users, aiming to provide a balance between noisy conditions and quiet conditions, as well as a bridge between research and real-world deployment. We also perform cross-dataset benchmarking experiments on existing methods and show that cross-dataset generalization remains challenging.

    Having obtained the dataset, we develop a new method that takes into consideration the sparse nature of kinship signals and makes use of the attention mechanism \cite{transformer,apps_of_transformer} to encode audio. While attention has been previously applied to feature fusion across multiple modalities for kinship verification \cite{talkin_dataset}, to the best of our knowledge, we are the first to propose such an approach for audio-based kinship verification. Existing literature on audio-based kinship verification \cite{kanav_dataset, talkin_family_dataset} rely either on traditional approaches such as GMM-UBM \cite{gmm-ubm} or purely convolutional architectures. To tackle the issue of training instability, we use a two-stage training scheme where we first start with the easier task of speaker recognition. In addition, existing literature on audio-based kinship verification rarely perform explainability analysis, which we will explore in this paper.
    
    Hence, the main contributions of our paper are the following:
	\begin{enumerate}
		\item We introduce ARKIN, a new uncontrolled speech dataset focusing on everyday recording conditions for audio-based kinship verification, and perform a series of baseline experiments on it.
		\item We propose CONVTRAP-TN, a novel transformer-based architecture for audio-based kinship verification. We further investigate CONVTRAP by analyzing model attention weights and conducting an ablation study.
        \item We investigate the cross-dataset robustness of existing methods for audio-based kinship verification.
	\end{enumerate}

\section{Related work}
	\begin{table*}[t]
		\centering
		\caption{Comparison of existing audio kinship datasets}
		\label{tab:dataset_comparison}
		\setlength{\tabcolsep}{6pt}
		\renewcommand{\arraystretch}{1.1}
		\begin{tabular}{lrrrcccc}
			\toprule
			\textbf{Dataset}&\textbf{\#Families}&\textbf{\#Speakers}&\textbf{\#Clips}&\textbf{Recording Condition}&\textbf{Language}&\textbf{Age Range}&\textbf{Availability}\\
			\midrule
			KAN-AV & 255 & 970 & 28\,003 & In-the-wild & English & 3--100 & On request\\
			TALKIN-Family & 246 & 1012 & 4048 & Semi-controlled & Mandarin & 5--81 & On request/Restricted\\
			FIW-MM & 150 & 500 & 937 & In-the-wild & English & N/A & On request/Restricted\\
			ARKIN (ours) & 86 & 256 & 2554 & \textbf{Uncontrolled} & Mandarin & 5--84 & On request\\
			\bottomrule
		\end{tabular}
	\end{table*}
	Few datasets contain both audio data as well as kinship labels. Current public datasets include the Kinship Age geNder - Audio Visual (KAN-AV) dataset \cite{kanav_dataset}, the TALking KINship - Family (TALKIN-Family) dataset \cite{talkin_family_dataset}, and the Families In Wild Multimedia (FIW-MM) database \cite{fiw-mm_dataset}. A comparison of the datasets (including ours) is presented in Table~\ref{tab:dataset_comparison}.

	KAN-AV \cite{kanav_dataset} is a manually annotated dataset consisting of in-the-wild audio and video data from the internet. The dataset contains around 28\,000 utterances, with roughly 66\% of the speakers having at least one kinship relation with another in the dataset. The majority of speakers included in the KAN-AV dataset are English-speaking public figures. KAN-AV is the largest publicly available audio dataset with kinship labels.

	TALKIN-Family \cite{talkin_family_dataset} is a manually annotated dataset collected from offline participants in a semi-controlled setting, where participants were instructed to record in a quiet location. The dataset contains 1012 subjects, totaling 4048 utterances. All speakers are from China and speak Mandarin, each having at least one kinship relation with another.
	
	FIW-MM extends the Families-In-Wild \cite{fiw_dataset} image dataset to include audio data. Similar to KAN-AV, speakers consist of public figures, the majority of which is English speaking and data is collected from the internet. The dataset contains 937 utterances (given by ``\# clips'' in the paper).
	
    We stress that our dataset is the first to provide Mandarin recordings in uncontrolled conditions using consumer-grade devices. We also differentiate our approach from sourcing recordings in-the-wild in that we do not source public recordings from the internet, which often have very low signal-to-noise ratio and are recorded on specialized equipment not available to the average population.

\section{ARKIN dataset}
	
	\subsection{Overview}
	The ARKIN dataset is a benchmark dataset that captures real-world recording conditions and serves as a bridge between datasets collected under controlled conditions and datasets with in-the-wild data. The dataset contains 256 speakers, each labeled with age, gender, and nationality. There are 2554 utterances, out of which 95\% are in Mandarin and only 5\% in English. We decided to primarily focus on Mandarin speech due to the current imbalance between the amount of English and non-English data available. Each utterance is provided with a manual text transcription and a language label. There are 230 kinship pairs, and every speaker has at least one kinship relation with another speaker. The dataset is organized by speaker IDs. Audio files are provided in 16-bit PCM format sampled at 48\,kHz.
	
	\subsection{Data collection and processing}
	
	In order to obtain audio with realistic environmental conditions that reflect average use-cases, we asked participants to record on any device they normally use (usually, this is their mobile phone) with no restriction on background conditions. Participants can record wherever they want, however they want, and whenever they want. As a result, we obtain a dataset with a large variety of background conditions.
	
	For the content of the recordings, participants were asked to read out first a fixed phrase, followed by an improvisation of anything they would like. Some participants did not improvise, whereas others had trouble improvising and asked for a reference instead, in which case we provided them with additional passages as inspiration.
	
\begin{figure}
		\centering
		\includegraphics[width=0.7\linewidth]{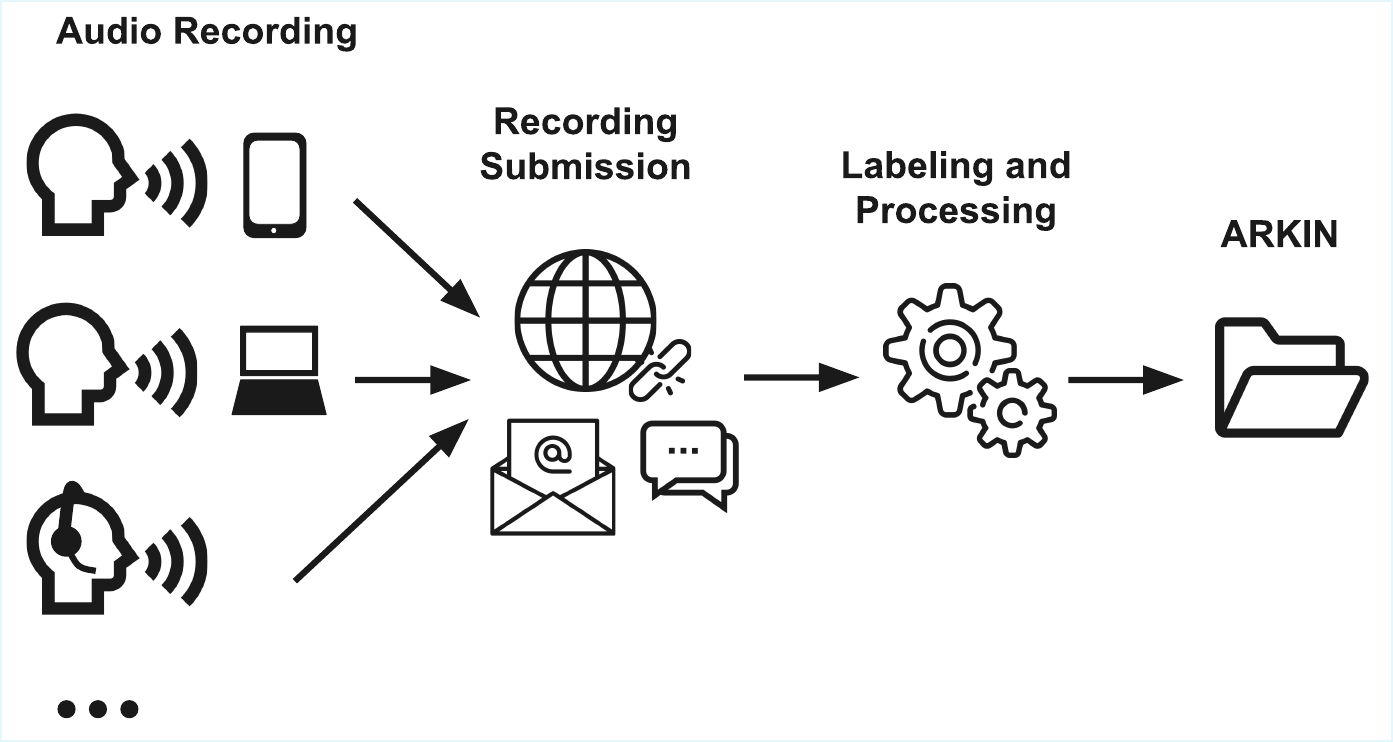}
		\caption{The data collection process}
		\label{fig:data_collection}
	\end{figure}

	We manually divide audio clips into utterances of varying lengths (roughly 0.3\,s\,-\,13.3\,s) and cut out sections that do not contain transcribable audio from the subject. We were able to capture a variety of speech patterns such as rapid pausing, false starts, and filler sounds. Because of this, our utterances can vary greatly in length and content. The data collection process is illustrated in Figure~\ref{fig:data_collection}.
	
	\section{Baseline experiments}
	\label{sec:guidelines}
	\subsection{Speech representation}
	To verify the performance of the ARKIN dataset, we need to extract speech features from utterances. We first resample our audio to 16\,kHz and extract MFCCs \cite{mfcc} with 24 coefficients using the Kaldi toolkit \cite{kaldi}. We choose a frame length of 25\,ms and a frame shift of 15\,ms. Then We apply Cepstral Mean and Variance Normalization (CMVN) and Voice Activity Detection (VAD) with an energy threshold of 5 to prepare the input features for subsequent modeling.

    We evaluate a series of baseline representations derived from these features. First, we use Kaldi to train a 512-component GMM-UBM \cite{gmm-ubm}, projecting the UBM statistics into a low-dimensional total variability space to extract 400-dimensional \textbf{i-vectors}\cite{ivector}. Second, we implement TDNN architectures \cite{xvector,ecapa-tdnn} trained from scratch on our dataset for speaker classification. From this model, we extract standard 512-dimensional \textbf{x-vectors}, reduced 300-dimensional embeddings obtained by modifying the final fully-connected layers to suit our data scale, as well as 256-dimensional embeddings from the \textbf{ECAPA-TDNN} architecture. Finally, to leverage transfer learning, we utilize a \textbf{ResNet} model \cite{speechbrain_resnet} from SpeechBrain \cite{speechbrain}. Pretrained on the massive VoxCeleb corpus \cite{voxceleb, voxceleb2}, this model yields compact 256-dimensional embeddings.

	We perform a 7:1:2 split to divide our data into disjoint training, validation, and test sets, ensuring no speakers in different sets share a kinship relation. For each utterance, we extract MFCC sequences, as well as i-vectors, x-vectors, and ResNet feature vectors, and perform, on the extracted features, various experiments described below. We average our results over five independent experiments. Details of the partitions are presented in Table \ref{tab:dataset_stats}.

    \begin{table*}[t]
  \caption{Detailed statistics of the ARKIN dataset partitions. \textbf{M}: Male, \textbf{F}: Female. Kinship types: \textbf{F-S} (Father-Son), \textbf{F-D} (Father-Daughter), \textbf{M-S} (Mother-Son), \textbf{M-D} (Mother-Daughter), \textbf{B-B} (Brother-Brother), \textbf{S-S} (Sister-Sister), \textbf{B-S} (Brother-Sister).}
  \label{tab:dataset_stats}
  \centering
  \resizebox{\textwidth}{!}{%
  \begin{tabular}{l|rrr|r|cc|rrrrrrr|r}
    \toprule
    \multirow{2}{*}{\textbf{Partition}} & \multicolumn{3}{c|}{\textbf{Speakers}} & \textbf{Audio} & \multicolumn{2}{c|}{\textbf{Age (Years)}} & \multicolumn{7}{c|}{\textbf{Kinship Pairs Distribution}} & \multirow{2}{*}{\textbf{Total Pairs}} \\
    \cline{2-4} \cline{6-14}
     & \textbf{M} & \textbf{F} & \textbf{Total} & \textbf{\# Utts} & \textbf{Mean $\pm$ SD} & \textbf{Range} & \textbf{F-S} & \textbf{F-D} & \textbf{M-S} & \textbf{M-D} & \textbf{B-B} & \textbf{S-S} & \textbf{B-S} & \\
    \midrule
    \textbf{Train} & 98 & 98 & 196 & 1796 & 34.3 $\pm$ 18.9 & 5--84 & 27 & 22 & 44 & 33 & 12 & 12 & 18 & 168 \\
    \textbf{Val}   & 9 & 10 & 19 & 286 & 29.3 $\pm$ 17.5 & 8--61 & 2 & 3 & 6 & 6 & 1 & 3 & 7 & 28 \\
    \textbf{Test}  & 19 & 22 & 41 & 472 & 29.5 $\pm$ 15.5 & 6--57 & 5 & 3 & 10 & 9 & 3 & 1 & 3 & 34 \\
    \midrule
    \textbf{Total} & \textbf{126} & \textbf{130} & \textbf{256} & \textbf{2554} & \textbf{33.2 $\pm$ 18.3} & \textbf{5--84} & \textbf{34} & \textbf{28} & \textbf{60} & \textbf{48} & \textbf{16} & \textbf{16} & \textbf{28} & \textbf{230} \\
    \bottomrule
  \end{tabular}%
  }
\end{table*}

	\subsection{Kinship verification}
	\label{sec:kinship_verification}
	
	For (audio-based) kinship verification, we use the standard DNN-based triplet network (TripletNet) \cite{tripletnet}, as well as our own method, which we refer to as the CONVolution-TRansformer-Attention-Pool Triplet Network (CONVTRAP-TN), to perform deep metric learning.
	
	We use a simple random triplet mining strategy. We construct $N$ positive ordered pairs of form $(\vec v_a,\vec v_p)$ of utterances from speakers with a kinship relation, in which the first speaker is the anchor and the second is the positive sample. During training time, we select a random speaker from the training set that does not share a kinship relation with the anchor and use that speaker as a negative sample to form the triplet $(\vec v_a,\vec v_p,\vec v_n)$.
	
	\subsubsection{TripletNet}
	\label{sec:basic_tripletnet}
	
	The triplets are fed into the TripletNet, which uses a shared embedding function $f_\text{DNN}:\mathbb{R}^d\to\mathbb{R}^{\lfloor d/2\rfloor}$, parameterized by a 4-layer DNN with layer norm, leaky ReLU, and dropout to map each sample onto an embedding metric space, obtaining triplets of form $(\vec u_a,\vec u_p,\vec u_n) = (f_\text{DNN}(\vec v_a),f_\text{DNN}(\vec v_p),f_\text{DNN}(\vec v_n))$. The first layers consist of $d$ neurons, whereas the output layer consists of $\lfloor{d/2}\rfloor$ neurons, followed by a normalization operation so that all embeddings sit on a unit hypersphere. The model then calculates the Euclidean distance between the embeddings and outputs ordered pairs of form $(d_p,d_n)=(\lVert\vec u_a-\vec u_p \rVert,\lVert\vec u_a-\vec u_n\rVert)$. The goal then is to minimize the triplet loss function given by \eqref{eq:tripletnet_loss}, where $M\in\mathbb{R}^+$ is the margin. We set $M = 10$.

	\begin{equation}
		L(d_p,d_n) = \sum_{i=1}^N \mathrm{ReLU}(d_p^2 - d_n^2 + M)
		\label{eq:tripletnet_loss}
	\end{equation}
	
	\subsubsection{CONVTRAP-TN}
	\begin{figure}
		\centering
		\includegraphics[width=0.55\linewidth]{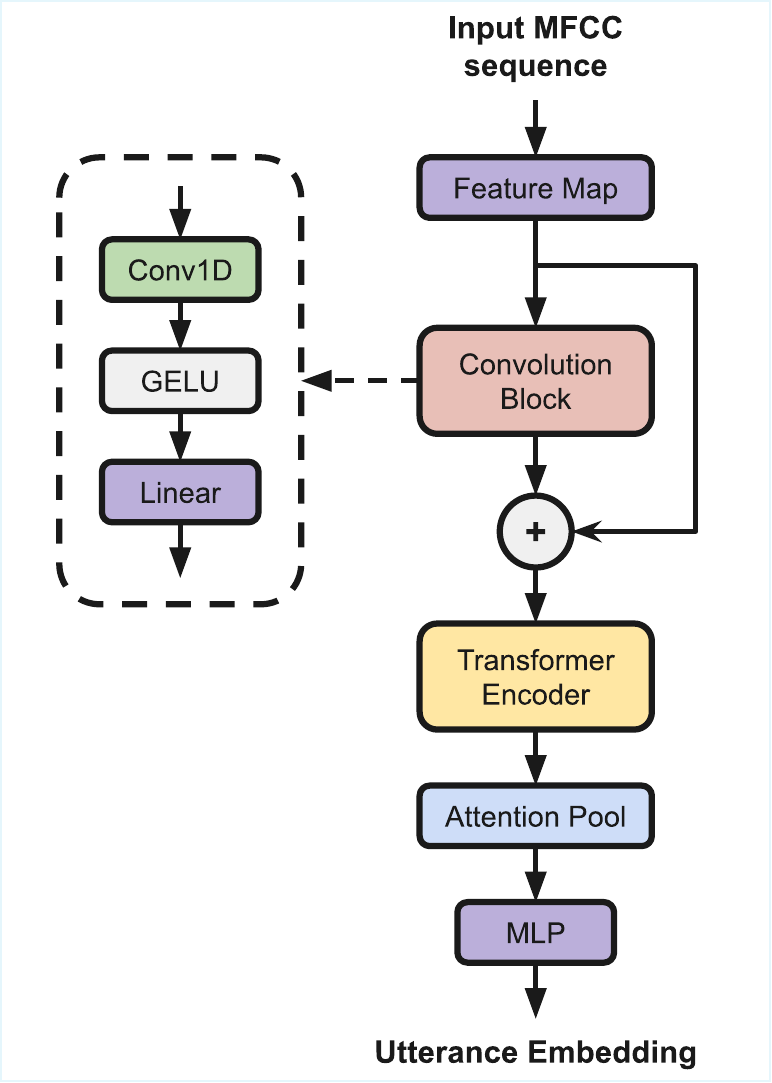}
		\caption{The CONVTRAP architecture}
		\label{fig:convtrap_tn_arch}
	\end{figure}
	
	Similarly, CONVTRAP-TN uses a shared embedding function to map triplets before computing the triplet loss. However, unlike the standard TripletNet, our CONVTRAP-TN uses the CONVTRAP embedding function (CNN + transformer + attention pooling). Our design takes into account the sparse nature of kinship signals. The CNN provides local context to the transformer, which then aggregates features that capture important clues spaced far apart, while discarding noise. The attention pooling module is specifically chosen to tackle the issue of kinship signal sparsity, as it can selectively aggregate features instead of diluting them, as other approaches such as mean pooling would do.
	
	CONVTRAP first takes in an MFCC sequence $C \in \mathbb{R}^{L\times 24}$ and projects it onto a high-dimensional space via a 3-layer DNN. We pick a hidden dimension of 512, following x-vectors. We then use a convolutional block (1D-Conv, GELU, Linear) as a form of positional encoding before feeding the sequence into a transformer encoder \cite{transformer} (3-layers and 8 heads empirically found to work best). The output is then using simple attention pooling \cite{attn_pool} defined as:
	
	\begin{equation}
		\mathrm{attnPool}_{\vec q}(C) = C^\mathsf T\mathrm{softmax}\left[\frac{\operatorname{mlp}(C)\vec q}{\sqrt{512}}\right],
		\label{eq:attention_pool}
	\end{equation}
	
	\noindent where $\vec q$ is a learnable query vector. Finally, the pooled vector is fed into a 2-layer DNN with layer norm, leaky ReLU, and dropout of 0.3. The complete CONVTRAP embedding function $f_\text{CONVTRAP}$ is visualized in Figure \ref{fig:convtrap_tn_arch}.

    Additionally, we employ a two-stage training approach to help with training stability. We first train the model on speaker recognition via a separate classification head (2-layer DNN with layer norm, leaky ReLU, and a dropout of 0.5) attached after the attention pool. We then freeze all parameters prior to the attention pool and train the partially frozen model on kinship verification. We are essentially extracting intermediate features, similar to x-vectors \cite{xvector}, with the difference being that we have a sequence instead of a single vector.

	\begin{table}
		\caption{Kinship verification balanced accuracy (\%, higher is better) at FAR = FRR, EER (\%, lower is better), and MinDCF with $P_\text{target}=0.5,C_\text{fn}=C_\text{fp}=1$ (lower is better)}
		\label{tab:kinship_verification_scores}
		\centering
		\setlength{\tabcolsep}{2pt}
		\begin{tabular}{llrrrrr}
			\toprule
			\textbf{Method}&\textbf{Features}&\textbf{Acc} $\uparrow$&\textbf{EER} $\downarrow$&\textbf{MinDCF} $\downarrow$\\
			\midrule
			TripletNet&i-vector&56.1&43.9&0.90\\
			\ &x-vector&63.4&36.6&0.73\\
			\ &x-vector 300-dim&64.7&35.3&0.69\\
            \ &x-vector ECAPA&55.2&44.8&0.87\\
			\ &ResNet&61.2&38.8&0.73\\
			COVTRAP-TN&MFCCs&\textbf{69.9}&\textbf{30.1}&\textbf{0.53}\\
			\bottomrule
		\end{tabular}
	\end{table}
    
	\begin{table*}
		\caption{Sensitivity (\%, higher is better) at FAR = FRR to different kinship types: \textbf{F-S} (Father-Son), \textbf{F-D} (Father-Daughter), \textbf{M-S} (Mother-Son), \textbf{M-D} (Mother-Daughter), \textbf{B-B} (Brother-Brother), \textbf{S-S} (Sister-Sister), \textbf{B-S} (Brother-Sister). Bold: most sensitive, underline: below average sensitive.}
		\label{tab:kinship_verification_scores_by_kin_type}
		\centering
		\setlength{\tabcolsep}{12pt}
		\begin{tabular}{llrrrrrrrr}
			\toprule
			\textbf{Method}&\textbf{Features}&\textbf{F-S}&\textbf{F-D}&\textbf{M-S}&\textbf{M-D}&\textbf{B-B}&\textbf{S-S}&\textbf{B-S}&\textbf{Average $\uparrow$}\\
			\midrule
			TripletNet&i-vector&\underline{42.8}&\underline{43.6}&81.1&67.1&71.8&\underline{38.2}&\textbf{87.2}&61.7\\
			\ &x-vector&\underline{27.8}&\underline{15.1}&75.5&73.3&\underline{21.9}&59.8&\textbf{88.8}&51.7\\
			\ &x-vector 300-dim&\underline{28.6}&57.8&79.5&63.7&\underline{25.0}&\underline{50.1}&\textbf{85.3}&55.7\\
            \ &x-vector ECAPA&\underline{31.0}&\underline{55.9}&\textbf{76.1}&70.4&64.9&\underline{48.2}&72.5&59.9\\
			\ &ResNet&\underline{32.8}&55.7&\textbf{65.7}&61.7&\underline{47.8}&\underline{43.5}&\underline{49.2}&50.9\\
			COVTRAP-TN&MFCCs&87.8&93.0&\underline{54.4}&\underline{60.2}&\underline{64.6}&\textbf{94.7}&\underline{63.7}&74.1\\
			\bottomrule
		\end{tabular}
	\end{table*}
    
	\begin{table}[t]
		\caption{Cross-dataset (train on KAN-AV, evaluate on ARKIN) kinship verification balanced accuracy scores (\%) at FAR = FRR. ft: fine-tuned, TN: TripleNet, vec.: vector}
		\label{tab:kanav_to_own_kinship_verification_scores}
		\centering
		\setlength{\tabcolsep}{3pt}
		\begin{tabular}{llrrr}
			\toprule
			\textbf{Method}&\textbf{Features}&\textbf{KAN-AV}&\textbf{ARKIN}&\textbf{ARKIN-ft}\\
			\midrule
			TN&i-vec&51.7&49.7&\textbf{53.5}\\
			\ &x-vec&53.4&48.9&52.6\\
			\ &x-vec 300-dim&52.8&\textbf{50.2}&49.8\\
            \ &x-vec ECAPA&51.9&49.8&49.2\\
			\ &ResNet&51.2&46.0&47.6\\
			CONVTRAP-TN&MFCCs&\textbf{56.9}&50.0&44.1\\
			\bottomrule
		\end{tabular}
	\end{table}
	We implement both models in PyTorch, with random Gaussian initialization, and train with a batch size of 16 using SGD with a learning rate of 0.001 and momentum 0.9. We compute the verification threshold using the validation set (at FAR = FRR), fixing it for the test set.
	
	Results are shown in Table \ref{tab:kinship_verification_scores}. Out of the methods we experimented with, our new approach, CONVTRAP-TN, performed the best. This is likely because our frozen transformer backbone outputs a sequence of features with no reduction in dimensionality, meaning the downstream attention pool and DNN have access to relatively uncompressed features extracted directly from raw MFCCs. The attention mechanism in the transformer also helped to capture any nuanced long-range patterns, which would be much harder for a CNN-only architecture. For this task, we also experimented with both 512-dimensional and 300-dimensional x-vectors and found that the lower 300-dimensional features performed better, likely due to the size of the dataset.

    Furthermore, to investigate biases in verifying different kin types, we present sensitivity scores by kin type for each model in Table \ref{tab:kinship_verification_scores_by_kin_type}. We observe that the sensitivity variation between kinship types is large. The standard TripletNet is least sensitive to father-son (F-S), father-daughter (F-D), brother-brother (B-B), and sister-sister (S-S) relations, and usually most sensitive to mother-son (M-S), mother-daughter (M-D), and brother-sister (B-S) relations. On the other hand, CONVTRAP-TN is highly sensitive to F-S and F-D relations and less sensitive to M-S and M-D relations. While sensitivity by kin type is somewhat proportional to the amount of training data of the kin type, the model itself also appears to be an influential factor. Certain kin types may naturally be easier to verify, but we do not have enough evidence to support this claim.
	
	\subsection{Cross-dataset kinship verification}
	
	We additionally perform cross-dataset baseline kinship verification. We choose to experiment with the KAN-AV dataset, as it is the largest currently available dataset for our needs. We train baseline models on the KAN-AV dataset and compare evaluation performance on KAN-AV with performance on our own dataset. We also evaluate the effectiveness of fine-tuning the model using the first 20\% of the training speakers from our dataset (detailed in our repository).
    
    We use the same methods explained in Section \ref{sec:kinship_verification}. Results are shown in Table \ref{tab:kanav_to_own_kinship_verification_scores}, from which it is clear that baseline models trained on KAN-AV suffer drastic performance degradation when tested on our dataset. Although CONVTRAP-TN performed the best, like other methods, its performance worsened by a large factor on KAN-AV compared to our ARKIN. We suspect one reason is that KAN-AV is mostly in English, while our dataset is mostly in Mandarin. Another reason is that recording environments of speeches, interviews, or YouTube videos differ greatly from our everyday style environment.
    
    As seen in Table \ref{tab:kanav_to_own_kinship_verification_scores}, after fine-tuning on a portion of ARKIN, the performance of CONVTRAP-TN, like most other approaches, worsened (worse than a random model). We suspect this is because training on the noisy KANAV has effectively served as a poor initialization for the model to start learning ARKIN, as the datasets are so differently distributed, making transfer learning difficult.
    
\begin{figure}
		\centering
		\includegraphics[width=0.6\linewidth]{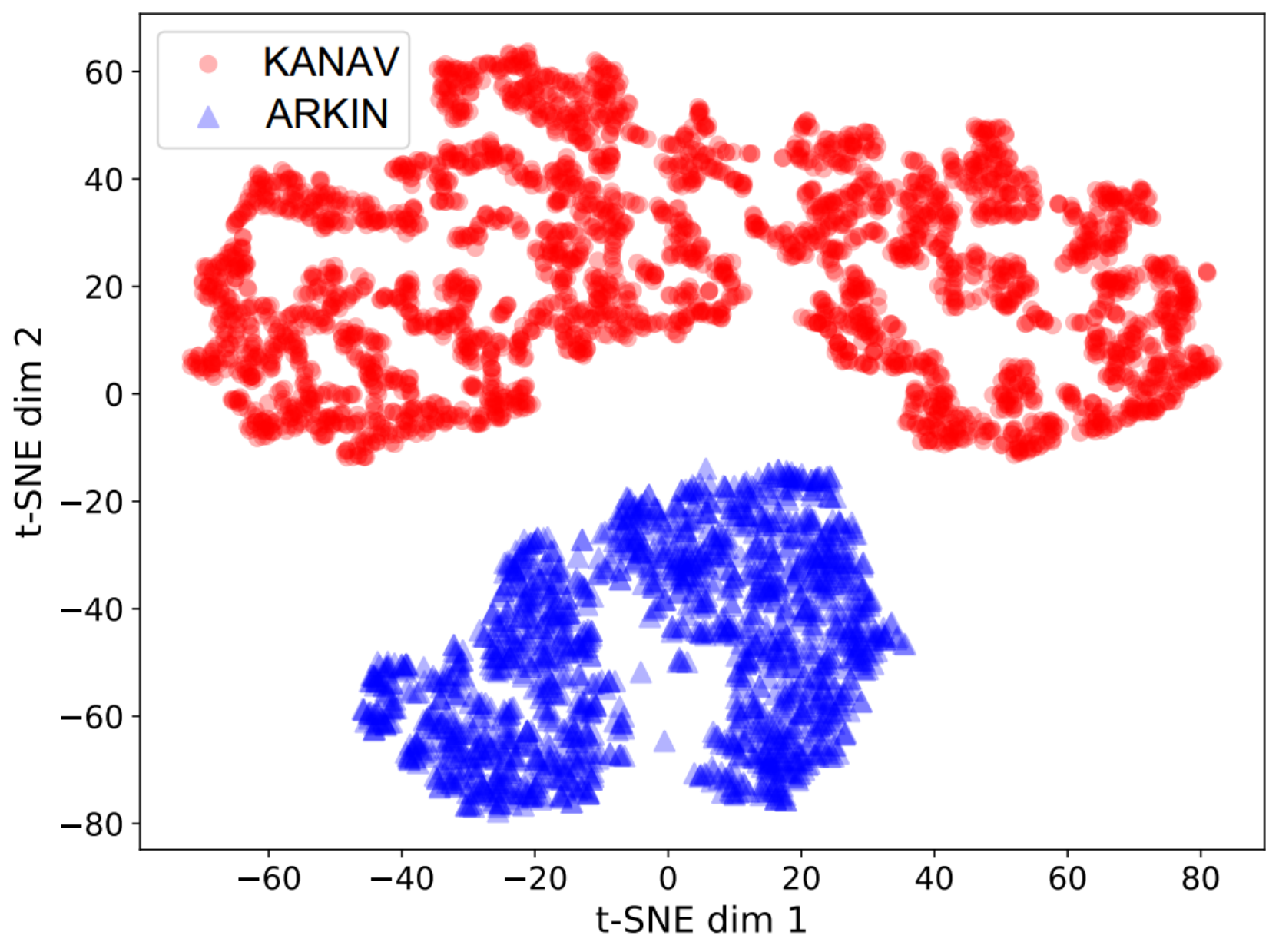}
		\caption{Embeddings (after dimensionality reduction) of utterances from KAN-AV (red circles) and ARKIN (blue triangles) obtained from a TripletNet trained on x-vectors}
		\label{fig:feat_vis}
	\end{figure}

    To clearly visualize the domain gap between our dataset and KAN-AV, we reduce high-dimensional kin embeddings (generated from 512-dimensional x-vectors) to 2-dimensional points using the t-distributed stochastic neighbor embedding (t-SNE) method \cite{tsne}. The points are plotted in Figure \ref{fig:feat_vis}, from which it is evident that there is a large domain gap between KAN-AV and our dataset.
    
	\subsection{Auxiliary Tasks}
	Given the rich meta-labels provided in ARKIN, we also perform baseline experiments on auxiliary tasks to validate the dataset's utility across different speech dimensions. We summarize the performance of these tasks in Table \ref{tab:all_benchmarks}.
	
	First, we evaluated identity-related tasks: speaker verification and speaker recognition. Despite the uncontrolled recording environments, the ResNet-based model achieved a recognition accuracy of 96.7\% and a verification accuracy of 82.5\%, outperforming traditional i-vector baselines. This confirms that the dataset preserves high-fidelity speaker identity features necessary for robust biometric modeling.
	
	Second, for paralinguistic and linguistic analysis, we benchmarked age estimation and Automatic Speech Recognition (ASR). In age estimation, ResNet yielded the lowest Mean Absolute Error (MAE) of 5.6 years. For ASR, the Whisper model demonstrated superior robustness against environmental noise with a Character Error Rate (CER) of 18.6\%, compared to 50.8\% for Wav2Vec 2.0. These results highlight the challenging nature of the dataset for standard acoustic models while showing that robust transcription is achievable.
	
	For conciseness, we only present a summary of results here, verifying the reliability of dataset labels. Detailed experimental setups, hyperparameters, and full reproduction scripts for all auxiliary tasks will be available in our public repository.

	\begin{table}[t]
	\caption{Baseline performance on all auxiliary tasks. Spk Ver./Rec.\ are accuracy (\%, higher is better), Age is MAE (years, lower is better), and ASR is CER (\%, lower is better).}
	\label{tab:all_benchmarks}
	\centering
	\resizebox{\columnwidth}{!}{
	\begin{tabular}{l|crrr}
		\toprule
		\multirow{2}{*}{\textbf{Method}} & \textbf{Spk Ver.} & \textbf{Spk Rec.} & \textbf{Age Est.} & \textbf{ASR} \\
		& \textbf{(Acc \%)} $\uparrow$ & \textbf{(Acc \%)} $\uparrow$ & \textbf{(MAE)} $\downarrow$ & \textbf{(CER \%) $\downarrow$} \\
		\midrule
		i-vector & 69.0 & 34.7 & 10.3 & - \\
		x-vector & 74.4 & 73.2 & 8.2 & - \\
		ResNet & \textbf{82.5} & \textbf{96.7} & \textbf{5.6} & - \\
		\midrule
		Wav2Vec 2.0 & - & - & - & 50.8 \\
		Whisper & - & - & - & \textbf{18.6} \\
		\bottomrule
	\end{tabular}%
	}
    \end{table}

\section{Further results on CONVTRAP-TN}
    \subsection{Attention visualization}

\begin{figure}[t]
        \centering
        \includegraphics[width=\linewidth]{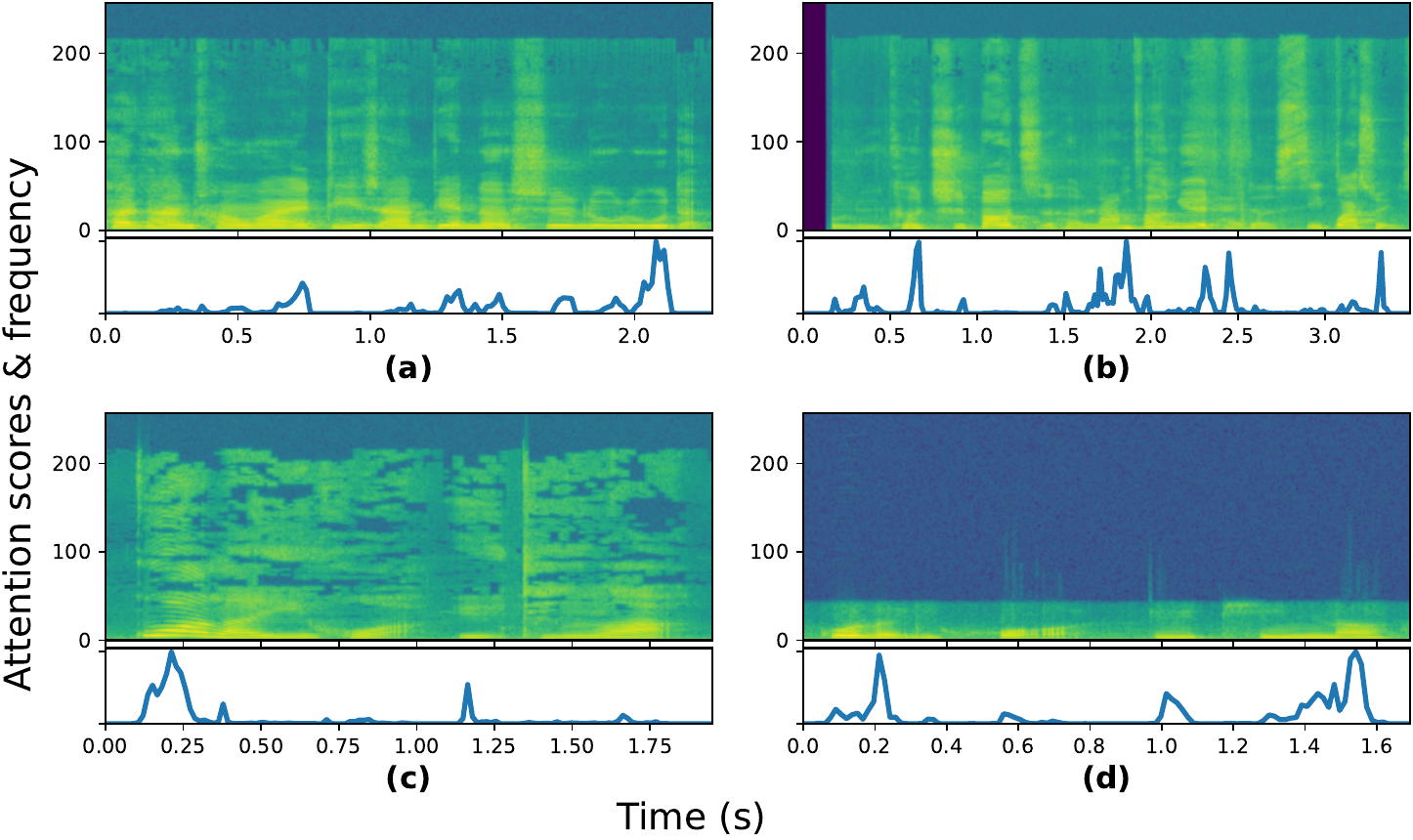}
        \caption{Visualization of input spectrograms alongside attention pool scores: (a) and (b) are a true positive pair, (b) and (c) are a true negative pair, (c) and (d) are a false positive pair.}
        \label{fig:attn_vis}
    \end{figure}

    We visualize attention scores from the attention pool component of the CONVTRAP model trained on ARKIN alongside input spectrograms. We pick TP, TN, and FP pairs and present them in Figure \ref{fig:attn_vis}. Results show that CONVTRAP consistently aggregates information from salient regions. In the failure (FP) case, the model still attended to active speech regions, meaning the error likely comes from ambiguity within learned representations rather than background confounding factors.

\subsection{Ablation study}

    To investigate the effectiveness of different components of CONVTRAP, we remove or decrease the capacity of different components and evaluate each ablated model. ``Two-stage" notes the full model trained using the two-stage approach described previously. ``Single layer" denotes the full model except with 1 transformer layer instead of 3. ``W/o convBlock" denotes the full model without the convolution block before the transformer encoder. ``W/o attnPool" denotes the full model with mean pooling instead of attention pooling. We also attempt to train the model directly on kinship verification without first training on speaker recognition, which we denote by ``Single-stage". We present results in Table \ref{tab:ablation_results}. 
    
    It is clear that the most vital aspects of our approach are attention pooling and two-stage training, as the model fails catastrophically without either of them. We suspect the reason why attention pooling performs much better than mean pooling is that our model does not have any other downsampling module. Since useful kinship clues are usually concentrated in just a few regions, mean pooling struggles to provide a useful aggregation, whereas attention pooling can learn to select the most relevant frames, as evidenced by Figure \ref{fig:attn_vis}. When it comes to single-stage training, kinship verification has proven to be too difficult a task. Without training the model first on speaker recognition, the model easily gets stuck in a bad local minimum due to the instability of the triplet network, unable to learn a good general representation of kinship. The convolution block also proved critical, which is not surprising considering the permutation-invariant nature of attention mechanisms. Lastly, having just a single transformer layer decreased performance by only 2.8\%, which is small compared to other components, though we suspect this may change if much larger datasets were used for training.

	\begin{table}[t]
		\caption{Kinship verification performance of ablated models trained on ARKIN. Balanced accuracy scores (\%, higher is better) at FAR = FRR, EER (\%, lower is better), and MinDCF with $P_\text{target}=0.5,C_\text{fn}=C_\text{fp}=1$ (lower is better).}
		\label{tab:ablation_results}
		\centering
		\setlength{\tabcolsep}{10pt}
		\begin{tabular}{lrrr}
			\toprule
			\textbf{Model}&\textbf{Acc $\uparrow$}&\textbf{EER $\downarrow$}&\textbf{MinDCF $\downarrow$}\\
			\midrule
			w/o attnPool&47.2&52.8&0.76\\
			w/o convBlock&53.4&46.6&0.72\\
            Single layer&67.1&32.9&0.57\\
			Single-stage&50.0&50.0&0.98\\
			Two-stage&\textbf{69.9}&\textbf{30.1}&\textbf{0.53}\\
			\bottomrule
		\end{tabular}
        \vspace{-0.5em}
	\end{table}

\section{Conclusion}

	We proposed CONVTRAP-TN, a new transformer-based architecture for kinship verification. Ablation study showed that attention pooling and two-stage training are the most influential components of the effectiveness of the approach. Through splitting kinship verification sensitivity scores by kinship type, we found that models are generally the most sensitive to brother-sister type relations and that the sensitivity variation between kinship types is large, raising concerns of fairness and bias, which is a direction for future improvements. We also introduced ARKIN, a kinship dataset enabling cross-dataset evaluation of audio models. Our evaluations reveal that models trained in one environment generalize poorly to others, regardless of dataset size, confirming a remarkable domain gap. One limitation is that we are only experimenting with supervised features. Future work will look at SSL features and focus on further developing robust audio models. To mitigate demographic homogeneity, we will expand data collection and evaluation across diverse populations.

\section*{Ethics declaration}
	
Approved by the institutional Research Ethics Committee of Imperial College London, the data collection protocol was fully compliant with GDPR and PIPL frameworks.
\section*{Disclosure of AI tools}
    No generative AI was used to produce any part of this paper.

\bibliographystyle{IEEEtran}
\bibliography{references}

\vspace{12pt}
\end{document}